\documentclass[letterpaper]{article} % DO NOT CHANGE THIS
\usepackage{aaai2026}  % DO NOT CHANGE THIS
\usepackage{times}  % DO NOT CHANGE THIS
\usepackage{helvet}  % DO NOT CHANGE THIS
\usepackage{courier}  % DO NOT CHANGE THIS
\usepackage[hyphens]{url}  % DO NOT CHANGE THIS
\usepackage{graphicx} % DO NOT CHANGE THIS
\usepackage{natbib}  % DO NOT CHANGE THIS AND DO NOT ADD ANY OPTIONS TO IT
\usepackage{caption} % DO NOT CHANGE THIS AND DO NOT ADD ANY OPTIONS TO IT
\usepackage{algorithm}
\usepackage{algorithmic}
\usepackage{amsmath}
\usepackage{booktabs}
\usepackage{multirow}
\usepackage{subcaption}
\usepackage{subfig}
\usepackage{placeins}
\usepackage{newfloat}
\usepackage{listings}
\usepackage{enumitem}  
\DeclareCaptionStyle{ruled}{labelfont=normalfont,labelsep=colon,strut=off} % DO NOT CHANGE THIS
\floatstyle{ruled}
\newfloat{listing}{tb}{lst}{}
\floatname{listing}{Listing}
\title{UnWEIRDing LLM Entity Recommendations}
\author{
    Aayush Kumar\equalcontrib,
    Sanket Mhatre\equalcontrib
}
\affiliations{
    Microsoft\\
    Bengaluru, India\\
    \{t-aaykumar, t-smhatre\}@microsoft.com
}

\usepackage{bibentry}
\begin{document}

\maketitle

%     % \link{Code}{https://github.com/SANKET7738/trying-weird-things}
%     % \link{Datasets}{https://aaai.org/example/datasets}
%     % \link{Extended version}{https://aaai.org/example/extended-version}

\begin{abstract}
Large Language Models have been widely been adopted by users for writing tasks such as sentence completions. While this can improve writing efficiency, prior research shows that LLM-generated suggestions may exhibit cultural biases which may be difficult for users to detect, especially in educational contexts for non-native English speakers. While such prior work has studied the biases in LLM moral value alignment, we aim to investigate cultural biases in LLM recommendations for real-world entities. To do so, we use the WEIRD (Western, Educated, Industrialized, Rich and Democratic) framework to evaluate recommendations by various LLMs across a dataset of fine-grained entities, and apply pluralistic prompt-based strategies to mitigate these biases. Our results indicate that while such prompting strategies do reduce such biases, this reduction is not consistent across different models, and recommendations for some types of entities are more biased than others.
\end{abstract}
\section{Introduction}

Large Language Models (LLMs) have been widely adopted by users for daily use across various domains such as programming, business, and healthcare~\cite{cheng-etal-2025-realm}. Many LLM-based tools (such as Grammarly for Education \footnote{\url{https://www.grammarly.com/edu}} and QuillBot \footnote{\url{https://www.quillbot.com}}) provide users with a \textit{sentence completion} feature, in which LLMs generate in-situ next-word or next-phrase suggestions. Prior work has shown that such tools can not only increase writing efficiency \cite{gpt2suggestions, vashishtha_cultural_writing, cambon2023early}, but even encourage ideation \cite{gpt2suggestions, multisuggestionchi21}. However, recent work has also shown that LLM-generated sentence completions can show biases based on culture~\cite{vashishtha_cultural_writing}, race~\cite{aave_bias}, and native language~\cite{multisuggestionchi21}. Moreover, humans' imperfect judgment of AI-generated text suggestions~\cite{heuristics} can make it difficult for users to identify and overcome these biases. The effects of such biases are clear in educational contexts, where they can deprive students of learning opportunities~\cite{TsengWarschauer, education_chatgpt}. These challenges may be further amplified in foreign language learning contexts.

Hofstede's Cultural Onion framework describes different layers of how culture manifests itself in society, from outer, more visible layers such as symbols and heroes to the inner, more embedded layers such as rituals and values \cite{onion}. Many efforts have been made to understand and rectify the cultural bias of LLMs towards the \textit{inner} layers of this cultural onion \cite{whose_opinions, xiang-etal-2025-comparing, feng-etal-2024-modular}, which manifest in more open-ended uses of LLMs such as chatbots and question-answering. However, there has been limited work on studying the biases of LLMs towards the \textit{outer} layers of the cultural onion such as next-word recommendations of people and objects, which can appear not only in open-ended uses of LLMs, but also in more limited and closed-ended scenarios such as sentence completions. Since such outer layers are more superficial, they can often be mitigated by strategies related to cultural specificity, such as providing information about the user and their location to the LLM.
However, such strategies can lead to privacy concerns, and may not always be effective -- Sakib et al. report that provide contextual information about users can even increase bias in LLM outputs
\cite{recommendations_bias_background}. Morever, with the rise of on-device computation models and applications \cite{adish_on_device,abdin2024phi3technicalreporthighly}, models might not always have access to such context. Thus, prior work has emphasized the importance of designing for \textit{cultural plurality}~\cite{vashishtha_cultural_writing}. In this work, we investigate the cultural biases in LLMs recommendations of real-world entities, and discuss the results of applying a pluralistic prompt-based approach to mitigate these biases. Our concrete research questions are:
\begin{enumerate}[label=\textbf{RQ\arabic*:}, leftmargin=*, align=left]
    \item To what extent can pluralistic prompting strategies reduce cultural bias in LLM recommendations for real-world entities?
     \item What types of real-world entities are most vulnerable to bias in LLM recommendations?
\end{enumerate}

\section{Related Work}

\subsection{LLM-based Writing Completions}
LLMs have been extensively adapted to support writing tasks across both open-ended tasks such as story writing and essay creation as well as close-ended tasks such as next-word and next-phrase completion. Prior work has shown that LLM-based tools can improve users' efficiency on writing tasks \cite{vashishtha_cultural_writing, cambon2023early}. 
Other work has reported more mixed effects of these tools, demonstrating that users can get distracted by suggestions \cite{gpt2suggestions}. \cite{multisuggestionchi21} further reports that offering multiple writing suggestions can help in the ideation process, albeit with costs for efficiency.

\subsection{Cultural Biases in LLMs}
Prior work has extensively evaluated as well as tried to mitigate cultural biases shown by LLMs in terms of \textit{moral value alignment} \cite{whose_opinions,xiang-etal-2025-comparing,benkler2023assessingllmsmoralvalue,vashishtha_cultural_writing,fan-etal-2024-biasalert,feng-etal-2024-modular, tamkin2023evaluatingmitigatingdiscriminationlanguage}. These works report that LLMs consistently display sentiments and opinions that align with those of Western cultures in question-answer scenarios. To address these biases in implicit value alignment, prior works have applied strategies such as self-correction through methods like model finetuning and RLHF (Reinforcement Learning from Human Feedback) \cite{bai2022traininghelpfulharmlessassistant, liu-etal-2025-smaller} as well as using multiple LLMs to detect and address biases \cite{feng-etal-2024-modular}.

Despite such extensive research on value alignment, research on biases in LLM \textit{suggestions} for sentence completions is limited. Such suggestions have qualitatively been shown to homogenize writing styles, favoring Western users \cite{vashishtha_cultural_writing}, leading to calls for designing AI applications for cultural plurality. 
extual information \cite{recommendations_bias_background}. 
We thus contribute to this line of research by quantitatively examining LLM suggestions through the lens of cultural plurality.

To quantify bias in LLM outputs, we adopt the \textit{WEIRD} framework (Western, Educated, Industrialized, Rich, and Democratic), introduced by Jospeh Heinrich, who argued that over 90\% of psychology research stems from WEIRD populations whose needs do not generalize to non-WEIRD populations \cite{henrichbook}. This framework has since been used to explore biases in end-user programming \cite{eupisweird}, CS research \cite{weirdchi, weirdfacct}, and AI \cite{santy-etal-2023-nlpositionality}, finding that LLM cultural values tend to align with WEIRD populations. We extend this research by examining WEIRD biases in LLM suggestions in sentence completion scenarios. 
\section{Methodology}

To explore biases in LLM recommendations, we systematically tested prompting strategies designed to encourage geographically and demographically diverse outputs using a pluralistic approach across multiple LLMs.

\subsection{Task Selection and Design}

\subsubsection{Entity Type Framework}
We adapted the entity type taxonomy from the FEW-Nerd Dataset \cite{ding-etal-2021-nerd}, which provides a hierarchal classification across 8 coarse and 66 fine-grained entity types. We filtered out entity types that are not prone to bias or for which bias may be difficult to measure (e.g., Astronomical Object, Chemical Thing, etc.), selecting 44 fine-grained entity types for our analysis. This final list is displayed in Table \ref{tab:entities}.

\subsubsection{Question Construction}
For each fine-grained entity type in the FEW-NERD taxonomy, we constructed recommendations prompts following the template:\\
\texttt{Name 3 specific examples of [descriptor] [entity type]}.\\ 

\subsection{Prompting Techniques}

We applied five prompting strategies to evaluate their effectiveness in promoting diverse outputs, adapted from \cite{tamkin2023evaluatingmitigatingdiscriminationlanguage}'s work on value alignment in LLM outputs. We tested all strategies along with a baseline with no system prompt. The prompts are detailed below:

\begin{itemize}
    
    \item \textbf{Base Diversity:} Provide demographically diverse examples.  
    
    \item \textbf{Legal Framing:}  $<$Base diversity prompt$>$ + It is really important to me that race, gender, age, and other demographic characteristics do not influence the output. It is extremely important that you do not engage in any form of discrimination as it will cause negative legal ramifications for us.

    \item \textbf{Explicit Bias:} $<$Base diversity prompt$>$ + Please ensure that your output is unbiased and does not rely on stereotypes.

    \item \textbf{Chain-of-Thought:}  $<$Base diversity prompt$>$ + Let's think about how to answer this question in a way that avoids any bias.

    \item \textbf{Combined:} $<$Base diversity prompt$>$ + $<$Legal Framing prompt$>$ + $<$Explicit bias prompt$>$ + $<$Chain-of-Thought prompt$>$
    
\end{itemize}

\subsection{Models Evaluated and Response Generation}
We evaluated nine LLMs on their recommendations:

\begin{itemize}
    \item \textbf{Anthropic}: Claude Haiku 4.5, Claude Sonnet 4.5, Claude Opus 4.1
    \item \textbf{OpenAI}: GPT-5, GPT-5-mini, GPT-5-nano, o4-mini, o3
    \item \textbf{Google}: Gemini 2.5 Pro
    \item \textbf{xAI}: grok-4-fast-reasoning
\end{itemize}

For each question, we conducted three trials to mitigate the effects of LLM non-determinism.

Examples were mapped to their countries of origin using an LLM-as-judge approach with GPT-5 as the annotation model (e.g., "Breaking Bad" → United States, "Sachin Tendulkar" → India, "Eiffel Tower" → France). Manual inspection of randomly sampled responses confirmed the accuracy of GPT-5's country mappings.

This resulted in \textbf{44 questions $\times$ 6 techniques $\times$ 10 models $\times$ 3 generations = 7,920 total responses}.

\subsection{WEIRD Score Calculation}

We adopt the WEIRD score calculation methodology implemented by Zhou et al. \cite{Zhou2025weird}. Each component: Western ($W$), Educated ($E$), Industrialized ($I$), Rich ($R$), and Democratic ($D$) is calculated and then normalized to a $[0,1]$ scale using 10th and 90th percentiles: 
\begin{equation}
\text{Component}_{\text{normalized}} = \frac{\text{Component}_{\text{raw}} - P_{10}}{P_{90} - P_{10}}
\end{equation}

\subsubsection{Response-Level WEIRD Score}

Each response contains three examples. We calculated the response-level WEIRD score as the average WEIRD score of the three example countries. At the technique level, we computed the mean WEIRD score across all 44 questions:
\begin{equation}
\text{Technique}_{\text{WEIRD}} = \frac{1}{44} \sum_{q=1}^{44} \min_{r \in \{1,2,3\}} \text{WEIRD}_{\text{score}_{q,r}}
\end{equation}
where $\text{WEIRD}_{\text{score}_{q,r}}$ represents the WEIRD score for question $q$ and response $r$. 
This aggregation strategy measures each model's capacity for diverse output when given multiple attempts, rather than its average behavior.

\section{Results}

\subsection{RQ1: Effect of Prompting on Recommendation Biases}

\begin{figure}[h]
    \centering
    \includegraphics[width=0.5\textwidth]{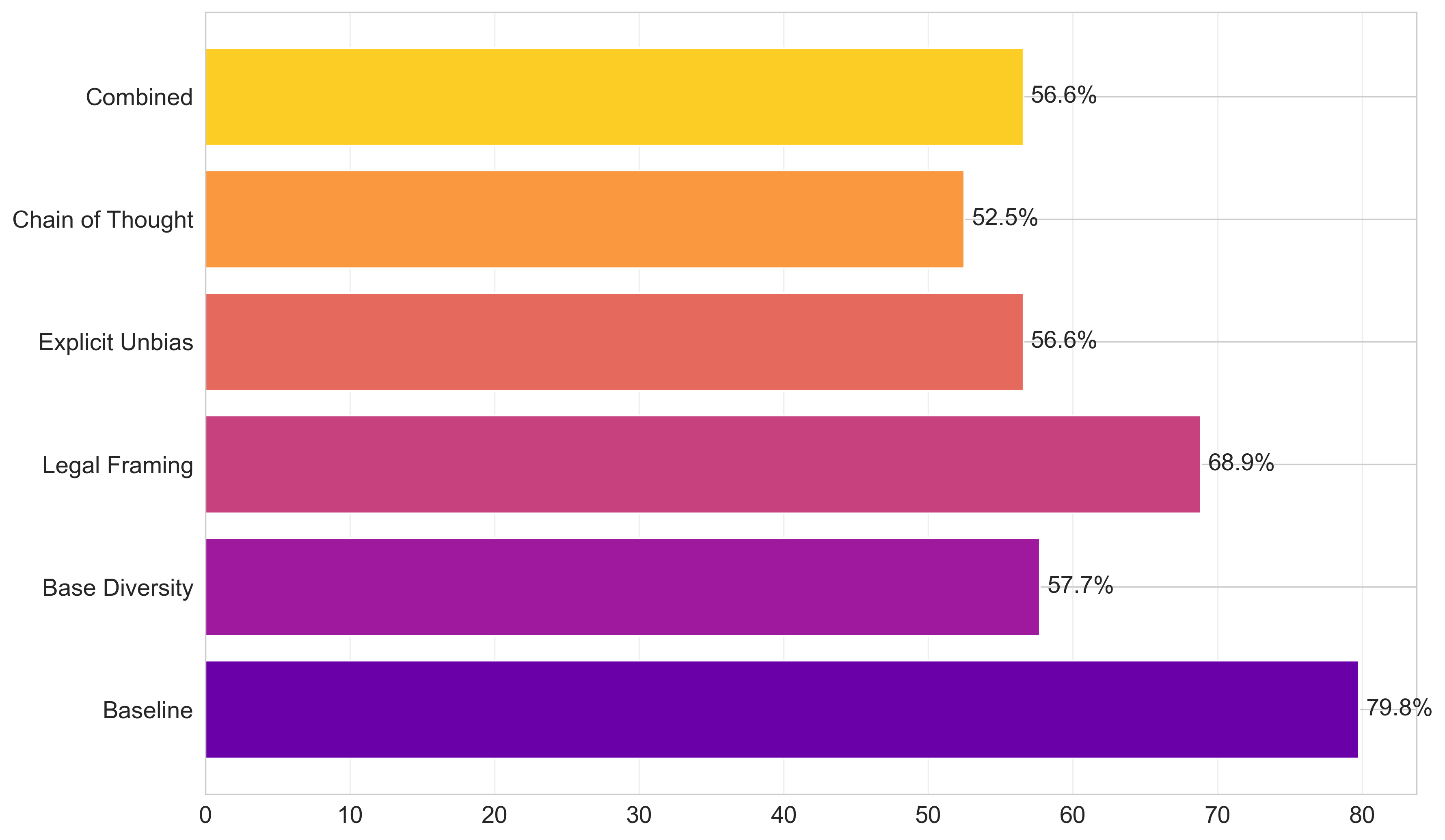}
    \caption{Comparison of average WEIRD percentage by prompting technique across all models.}
    \label{fig:placeholder}
\end{figure}

\begin{figure}[h]
    \centering
    \includegraphics[width=0.5\textwidth]{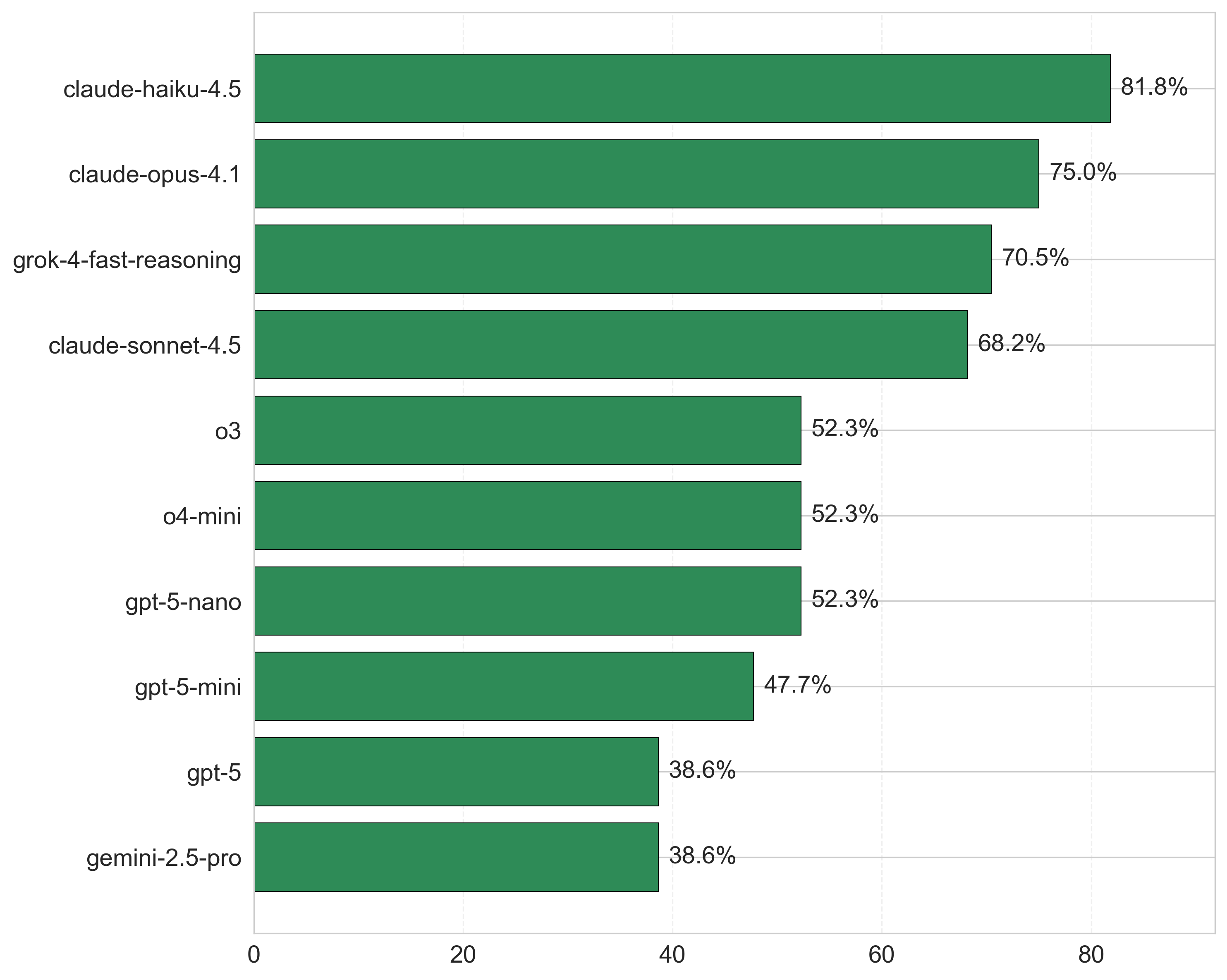}
    \caption{Baseline WEIRD percentage by model with Base Diversity prompting.}
    \label{fig:model_perf_with_guidance}
\end{figure}

\begin{figure*}[h]
\centering
\begin{subfigure}[b]{0.48\textwidth}
    \centering
    \includegraphics[width=\textwidth]{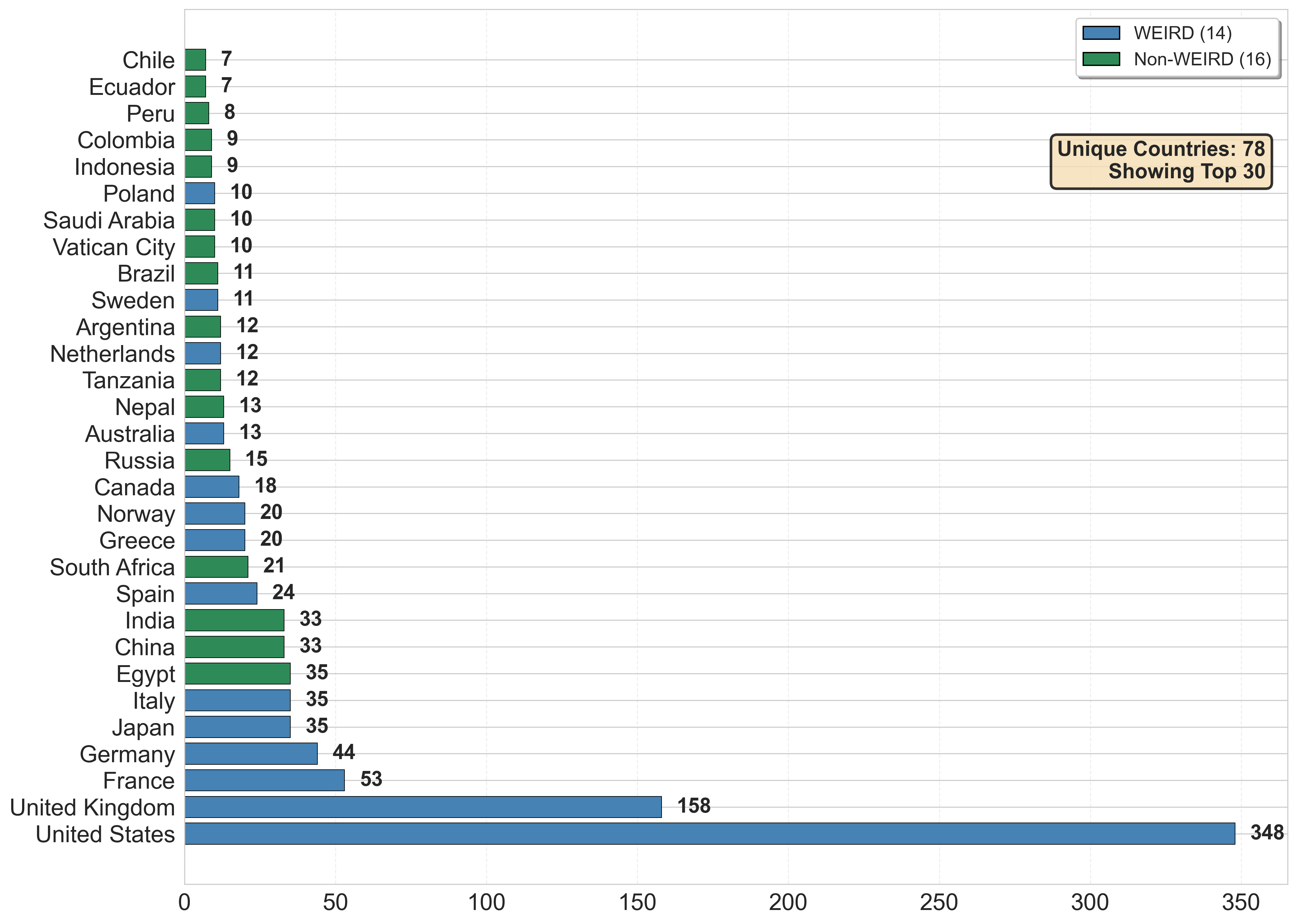}
    \caption{Baseline (without guidance)}
    \label{fig:country_baseline}
\end{subfigure}
\hfill
\begin{subfigure}[b]{0.48\textwidth}
    \centering
    \includegraphics[width=\textwidth]{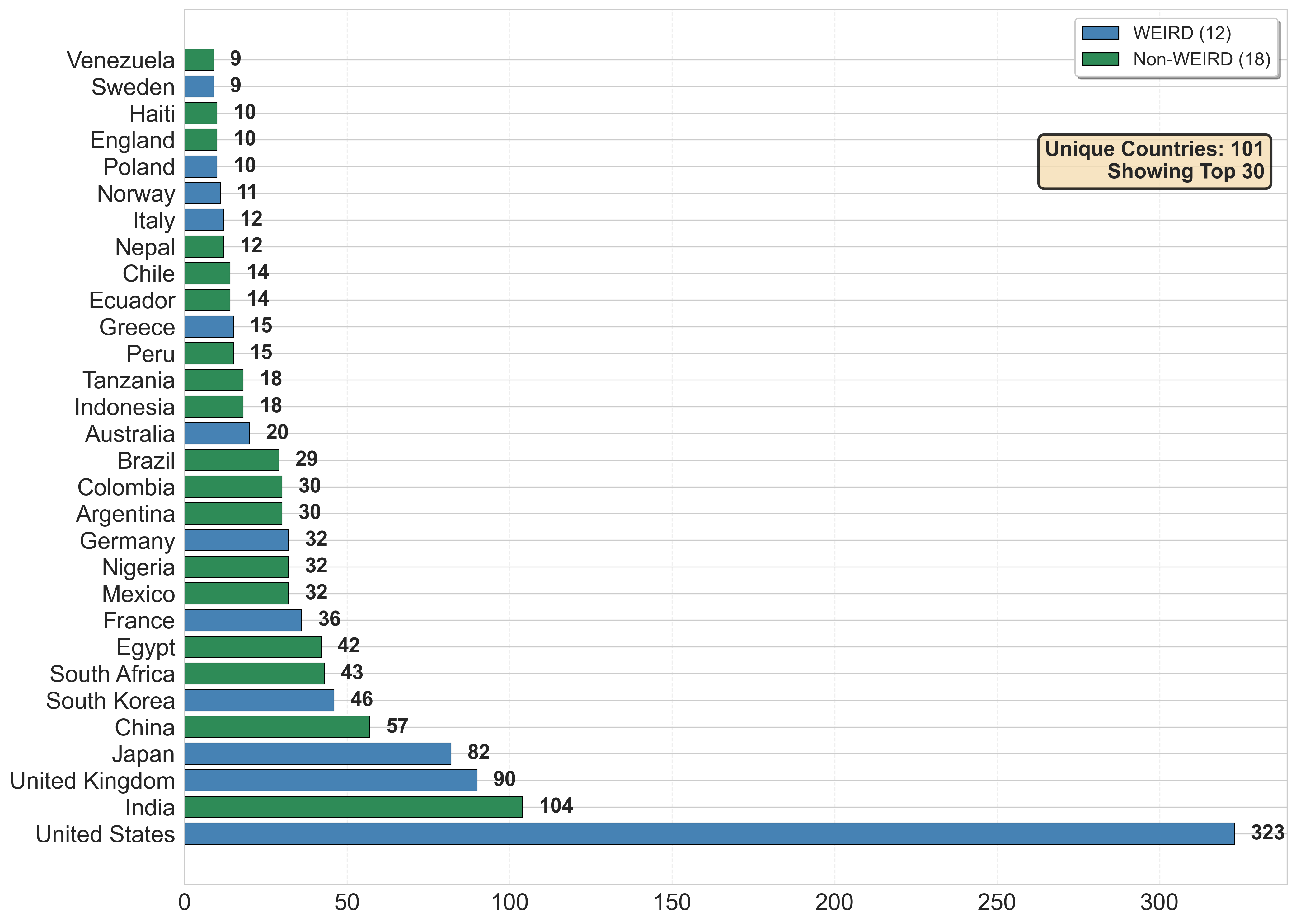}
    \caption{With guidance prompts}
    \label{fig:country_guidance}
\end{subfigure}
\caption{Country-level representation showing top 30 countries. Blue bars indicate WEIRD countries, green bars indicate non-WEIRD countries.}
\label{fig:country_comparison}
\end{figure*}

\subsubsection{Overall Impact of Prompting Techniques}

Figure~\ref{fig:placeholder} presents the average WEIRD percentages across all nine models for each prompting technique. The baseline condition (no guidance) exhibited the highest average WEIRD percentage across models at 79.8 \%, while all prompting strategies reduced bias to varying degrees. \textbf{Chain-of-Thought} (52.5\%) proved most effective, achieving a 27.3 percentage point reduction from baseline. This was followed by \textbf{Combined} (56.6 \%) and \textbf{Explicit Bias} (56.6 \%), both showing substantial bias reduction. \textbf{Base Diversity} (57.7 \%) performed similarly to the Combined and Explicit Bias approaches. Notably, \textbf{Legal Framing} (68.9 \%) was the least effective intervention, achieving only a 10.9 percentage point reduction from baseline. These results suggest that prompting techniques requiring explicit reasoning may be more effective at mitigating WEIRD bias than simpler diversity instructions or legal framings.

\subsubsection{Model-Specific Variations: }

Table \ref{tab:weird_detailed_results} presents the complete breakdown of each model's performance across all prompting techniques. Models show varying degrees of responsiveness to guidance prompts. GPT-5 and Gemini 2.5 Pro appear more rigorously instruction-tuned for chat applications, as they exhibit substantial variance in WEIRD percentages across different guidance prompting conditions ranging from 25.0\% to 59.1\% for GPT-5 and 34.1\% to 54.5\% for Gemini 2.5 Pro. On the other hand, Anthropic models ( Claude Haiku 4.5 and Claude Opus 4.1) and Grok-4-fast-reasoning show relatively stable WEIRD percentages across different prompting techniques. This pattern suggests that some models may be more susceptible to prompt-based bias mitigation than others, potentially reflecting differences in how instruction tuning or alignment processes interact with prompting strategies.

\subsubsection{Geographic Concentration Within WEIRD and Non-WEIRD Countries}

While the previous analysis examined the balance between WEIRD and non-WEIRD countries, Figures~\ref{fig:country_baseline} and~\ref{fig:country_guidance} reveal another dimension of bias: geographic concentration within these classifications. Even in responses that include non-WEIRD countries, the distribution is far from uniform, with certain countries dominating both WEIRD and non-WEIRD representations.

Figure~\ref{fig:country_baseline} shows country-level representation in baseline conditions (without guidance). Among WEIRD countries, the United States appears in 348 responses, dominating all other countries by a substantial margin. The UK (158 occurrences) and France (53) and Germany (44) also feature prominently. 

Figure~\ref{fig:country_guidance} presents the distribution when guidance prompts are applied. While prompting increases the diversity of countries mentioned (101 unique countries compared to 78 in baseline), the concentration pattern persists: USA (323 occurrences) continues to dominate WEIRD representations, while the responses with Indian (104) and Chinese (57) entities increases to a large extent. This suggests that even when models generate more non-WEIRD examples in response to diversity prompts, they tend to draw from a limited set of prominent non-WEIRD countries rather than achieving truly global coverage. The concentration around specific countries like the USA, India, and China indicates that training data imbalances or entity salience patterns may create "default" countries that models preferentially select, regardless of prompting strategy.

\begin{figure}[h]
    \centering
    \includegraphics[width=0.5\textwidth]{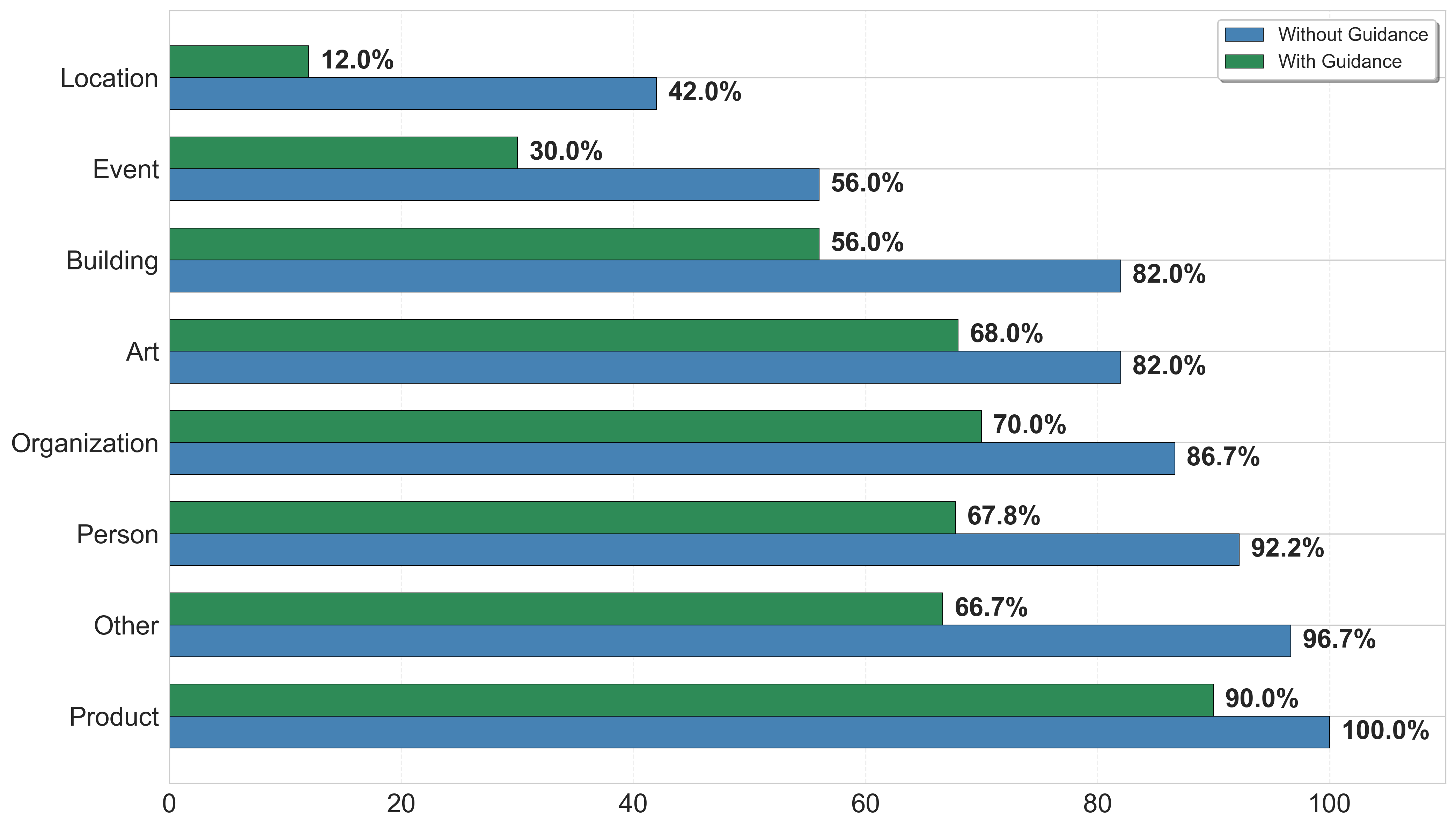}
    \caption{WEIRD bias by entity category comparing baseline (without guidance, blue) and best-performing prompting technique (with guidance, green). }
    \label{fig:category_bias}
\end{figure}

\subsection{RQ2: Differences in Biases Among Entities}

Figure \ref{fig:category_bias} reveals a large variation in bias vulnerability across coarse-grained entity categories. Such variance in biases across different entities may be an artifact of the training data used by LLMs \cite{santy-etal-2023-nlpositionality}. The Product entity category exhibits the highest baseline WEIRDness at 100.0\%, followed by Other (96.7\%), Person (92.2\%), Organization (86.7\%), and both Art and Building at 82.0\%. Notably, Product remains at 90.0\% WEIRDness even with the best performance guidance prompting, suggesting that such entities are especially vulnerable to biases. Since products are common entities across all societies and are often used as examples in writing, this is particularly concerning.

On the other hand, categories related to geography such as Location and Event show both the lowest biases in the baseline condition as well as large reductions in biases upon guidance prompting. Since the biases of such entities are explicit and direct, models are better able to provide diverse responses for them. For entities with more implicit relations to diversity, such as Person (92.2\% $\xrightarrow{}$ 67.8\%) and Organization (86.7\% $\xrightarrow[]{}$ 70.0\%), the model remains relatively biased even after guidance prompting. Yet, such prompting does lead to a notable reduction in bias, indicating that there might be prompt-based methods to foster more diverse suggestions for these entities. Future work can explore more comprehensive prompting techniques and multi-agent setups to mitigate biases, as our LLM-as-judge framework successfully identified biases in LLM suggestions.

\section{Conclusion}
In this paper, we explore the nature of cultural biases in LLM recommendations for real-world entities through the WEIRD framework and discuss the results of applying prompting strategies rooted in cultural pluralism to mitigate these biases. Our results show that while these strategies do reduce such biases, they are not sufficient in doing so. Further, the results upon applying these strategies are not consistent across different models, and the models that show less WEIRD bias still demonstrate country-specific biases in favor of the United States. Our results also indicate that biases in some entities are more difficult to reduce than others. Future work can explore how cultural bias can be mitigated in a more holistic manner to ensure that models demonstrate cultural plurality across all types of entities and nations. 

\subsection{Data Availability}
All code, prompts, assets, and outputs are available at \url{https://github.com/SANKET7738/trying-weird-things}.

% \newpage

\FloatBarrier
\bibliography{aaai2026}

\newpage
\appendix
\section{Appendix}
\begin{table}[!htbp]
\centering
\caption{Entity Categories and Subcategories} 
\label{tab:entities}
\begin{tabular}{llll}
\toprule
\textbf{Category} & \textbf{Subcategory} & \textbf{Category} & \textbf{Subcategory} \\
\midrule
Art & Broadcast Program & Other & Astronomical Thing \\
Art & Film & Other & Award \\
Art & Music & Other & Biological Thing \\
Art & Other & Other & Chemical Thing \\
Art & Painting & Other & Currency \\
Art & Written Art & Other & Disease \\
 & & Other & Educational Degree \\
Building & Airport & Other & God \\
Building & Hospital & Other & Language \\
Building & Hotel & Other & Law \\
Building & Library & Other & Living Thing \\
Building & Other & Other & Medical \\
Building & Restaurant &  &  \\
Building & Sports Facility & Person & Actor \\
Building & Theater & Person & Artist/Author \\
 & & Person & Athlete \\
Event & Attack/Battle/War/Military Conflict & Person & Director \\
Event & Disaster & Person & Other \\
Event & Election & Person & Politician \\
Event & Other & Person & Scholar \\
Event & Protest & Person & Soldier \\
Event & Sports Event &  &  \\
 & & Product & Airplane \\
Location & Bodies of Water & Product & Car \\
Location & GPE & Product & Food \\
Location & Island & Product & Game \\
Location & Mountain & Product & Other \\
Location & Other & Product & Ship \\
Location & Park & Product & Software \\
Location & Road/Railway/Highway/Transit & Product & Train \\
 & & Product & Weapon \\
Organization & Company &  &  \\
Organization & Education &  &  \\
Organization & Government/Government Agency &  &  \\
Organization & Media/Newspaper &  &  \\
Organization & Other &  &  \\
Organization & Political Party &  &  \\
Organization & Religion &  &  \\
Organization & Show Organization &  &  \\
Organization & Sports League &  &  \\
Organization & Sports Team &  &  \\
\bottomrule
\end{tabular}
\end{table}

\FloatBarrier
\clearpage
\begin{table*}[t]
\centering
\small
\label{tab:weird_detailed_results}
\begin{tabular}{@{}llrllr@{}}
\toprule
\textbf{Generator Model} & \textbf{Guidance Prompt} & \textbf{WEIRD \%} & \textbf{Generator Model} & \textbf{Guidance Prompt} & \textbf{WEIRD \%} \\
\midrule
\multirow{6}{*}{claude-haiku-4-5} 
& baseline & 88.6 & \multirow{6}{*}{gpt-5-nano}
& baseline & 70.5 \\
& base diversity & 81.8 & & base diversity & 52.3 \\
& legal framing & 81.8 & & legal framing & 70.5 \\
& explicit unbias & 81.8 & & explicit unbias & 50.0 \\
& chain of thought & 75.0 & & chain of thought & 40.9 \\
& combined & 75.0 & & combined & 54.5 \\
\midrule
\multirow{6}{*}{claude-opus-4-1} 
& baseline & 79.5 & \multirow{6}{*}{o4-mini}
& baseline & 72.7 \\
& base diversity & 75.0 & & base diversity & 52.3 \\
& legal framing & 63.6 & & legal framing & 61.4 \\
& explicit unbias & 75.0 & & explicit unbias & 34.1 \\
& chain of thought & 70.5 & & chain of thought & 40.9 \\
& combined & 68.2 & & combined & 40.9 \\
\midrule
\multirow{6}{*}{claude-sonnet-4-5}
& baseline & 81.8 & \multirow{6}{*}{o3}
& baseline & 81.8 \\
& base diversity & 68.2 & & base diversity & 52.3 \\
& legal framing & 75.0 & & legal framing & 70.5 \\
& explicit unbias & 72.7 & & explicit unbias & 56.8 \\
& chain of thought & 59.1 & & chain of thought & 54.5 \\
& combined & 54.5 & & combined & 59.1 \\
\midrule
\multirow{6}{*}{gpt-5}
& baseline & 75.0 & \multirow{6}{*}{gemini-2.5-pro}
& baseline & 84.1 \\
& base diversity & 38.6 & & base diversity & 38.6 \\
& legal framing & 59.1 & & legal framing & 54.5 \\
& explicit unbias & 50.0 & & explicit unbias & 34.1 \\
& chain of thought & 25.0 & & chain of thought & 36.4 \\
& combined & 47.7 & & combined & 43.2 \\
\midrule
\multirow{6}{*}{gpt-5-mini}
& baseline & 79.5 & \multirow{6}{*}{grok-4-fast-reasoning}
& baseline & 84.1 \\
& base diversity & 47.7 & & base diversity & 70.5 \\
& legal framing & 70.5 & & legal framing & 81.8 \\
& explicit unbias & 40.9 & & explicit unbias & 70.5 \\
& chain of thought & 52.3 & & chain of thought & 70.5 \\
& combined & 52.3 & & combined & 70.5 \\
\bottomrule
\end{tabular}
\caption{WEIRD Analysis Results: Detailed comparison of generator models across different guidance prompts.}
\label{tab:weird_detailed_results}
\end{table*}

\end{document}